\documentstyle[twocolumn,twoside]{article}
\newcommand{\hb}{\\ \hspace*{2ex}}

\begin{document}
\title{MAGNETIC FIELDS OF ISOLATED NEUTRON STARS: EVIDENCE FOR DECAY}
\author{S.B.\,Popov, M.E.\,Prokhorov\\[2mm]
 Sternberg Astronomical Institute, Moscow State University,\hb
 Universitetskii pr. 13, Moscow 119899 Russia,\hb
{\em polar@sai.msu.ru; mystery@sai.msu.ru}\\[2mm]
}
\date{}
\maketitle

ABSTRACT.
We show, that different types of isolated neutron stars (INSs)
show evidence in favor of magnetic field decay in these objects
and discuss how observations of INSs can help
to constrain models of field decay.
\\[1mm]

{\bf 1. Introduction}\\[1mm]

 Evolution of isolated stars of any type is much more simple
in comparison with the case when the object is a member of a close
binary system, where mass transfer takes place.
That is why looking for some undetected process it is much better to 
search for it in a more simple isolated case.

 In this short note we will speak about magnetic field decay (MFD)
in neutron stars (NSs) (see the paper by Konenkov and Geppert (2001) 
for recent calculations of MFD). Of course it is much more convenient to 
discuss this process in INSs, where there is no
influence of strong accretion from the second companion.

 In some details we will discuss anomalous X-ray pulsars (AXPs),
soft gamma repeaters (SGR) and ROSAT INSs (RINSs). See for recent 
reviews on SGRs (Hurley 2000), AXPs --- (Mereghetti 1999),
RINSs --- (Treves et al. 2000, Popov 2001). 
In the final section 
we briefly review indications of MFD in other types of INSs.\\[2mm] 

{\bf 2. Soft gamma repeaters and anomalous X-ray pulsars:
period clustering for magnetars}\\[1mm]

 Most probably AXP and SGR are ultramagnetized INSs,
so called {\it magnetars} (Duncan, Thompson 1992).
The alternative hypothesis that these objects are young INSs
accreting from a remnant (fall-back) disk meets  difficulties
(see Kaplan et al. 2001 and Duncan 2001
for a recent discussion on difficulties of both models).

  The main feature of SGR and AXP which we are going to discuss here is
period clustering: periods of all this objects are situated
in a very narrow range 5--12 s (see the table). 
This phenomenon can be easily explained by MFD.
We follow here Colpi et al. (2000).

  The authors discuss three main mechanisms of the MFD in magnetars
(in normal NSs mechanisms can be different):
ambipolar diffusion in the irrotaional and solenoidal modes and the Hall
cascade. For all three cases Colpi et al. calculate $p$---$\dot p$ diagrams.

 They obtain that if the Hall cascade is the main working mechanism and 
if typical decaying time scale is $\sim 10^4$ years, 
then it is possible to explain the observed data. 
For these parameters INSs reach asymptotic periods, i.e.
their spin rate is not changing significantly during their
subsequent evolution. The period "remembers"
the value it had before the magnetic field significantly decrease
as far as spin-down rate is strongly dependent on the value of the magnetic
field.

 Note, that as far as during MFD the NS's crust is heated,
this phenomenon is necessary to explain relatively strong 
thermal radiation of these sources.\\[2mm]

\begin{table}
\caption{Periods of INSs (from Mereghetti 1999, Hurley 2000,
Treves et al. 2000, Zampieri et al. 2001)}
\begin{tabular}{|l||c|c|}
Source                & Type & Period, s \\
                      &      &           \\ 
1E 1048-59.37         & AXP  &  6.44     \\
AX J1845-0258         & AXP  &  6.97     \\  
1E 2259+586           & AXP  &  6.98     \\
4U 0142+61.5          & AXP  &  8.69     \\
RX J1708.49-400.90    & AXP  & 11.00     \\
1E 1841-045           & AXP  & 11.77     \\
                      &      &           \\
SGR 1900+14           & SGR  &  5.16     \\
SGR 1627-41           & SGR  &  6.41     \\
SGR 1806-20           & SGR  &  7.48     \\
SGR 0526-66           & SGR  &  8.1      \\
                      &      &           \\
1RXS J130848.6+212708 & RINS &  5.15     \\
RX J0720.4-3125       & RINS &  8.37     \\
RX J0420.0-5022       & RINS & 22.7      \\
\end{tabular}
\end{table}

{\bf 3. Seven ROSAT INSs: old accretors or/and young coolers}\\[1mm]

 Now we know seven RINSs (Treves et al. 2000, Zampieri et al. 2001).
Their nature is not clear. They can be young cooling or old accreting INSs.
For three of them spin periods are determined (see the table).
In this section we will try to show, that in both alternative hypothesis
it is necessary to introduce MFD.\\[2mm]

{\it 3.1. Old accretors}\\[1mm]

  Old low velocity INSs can reach accretion stage (see a recent
review in Treves et al. 2000). In (Popov et al. 2000a,b) we calculated
evolution of populations of INSs in order to explore relative numbers of INSs
at different stages. One of the result is that without MFD it is
impossible to explain the observed population of RINSs by old accretors.

 Fitting parameters of decay it is possible to obtain necessary number of
accretors. It is unexplored yet if it is possible to explain the observed
log N -- log S distribution.

 After the first period determination for RINS was announced
(Haberl et al. 1996) two papers appeared, 
in which authors explained this period as 
a result of MFD in old accreting INS, (Wang 1997, Konenkov, Popov 1997).

 The idea is the following. For such short periods and typical parameters of
INSs and interstellar medium accretion is not aloud.
The only possibility is that the magnetic field is very low, $\sim 10^8$ G.
If the NS was born with such field and typical period of a young NS
(about 20 ms), then it is impossible to decrease spin period up to 8.4 s even
in $10^{10}$ years! So, the INS should be born with "normal" pulsar
parameters, and then magnetic field decays and the period again "remembers"
the value, when the field was strong.\\[2mm]

{\it 3.1.1 Constraints on magnetic field decay}\\[1mm]

  If we assume, that accreting INS are really observed, then we can put
limitations onto the models of MFD. 

 MFD can both increase and decrease number of accretors (Colpi et al. 1998,
Livio et al. 1998, Popov et al. 2000a). In (Popov, Prokhorov 2000) we 
tried to estimate these limitations for exponential decay,
$\mu=\mu_o \exp{-t/t_d}, \, \mu>\mu_b$.
In this case we can discuss two parameters: time scale, $t_d$, 
and bottom magnetic moment, $\mu_b$.
The later is the value, when decay stops. 

 For typical "pulsar" parameters of NSs we find out that intermediate
values of the bottom magnetic moment, $10^{28}\, {\rm G cm}^3<\mu_b <
10^{29.5}\, {\rm G cm}^3$, and  time scale, $10^7\, {\rm yrs}<t_d<10^8\,
{\rm yrs}$,  are forbidden.\\[2mm]

{\it 3.2. Young coolers}\\[1mm]    
 
  If we try to explain all observed RINSs as young cooling NSs,
then we come to a conclusion, that log N --- log S distribution
can not be explained without an assumption that the total number
of INSs is higher than the one derived from radiopulsar statistics
at least locally in time ($<10^7$ years) and space ($<300$ pc around the Sun)
or that the time, when an INS is still hot, is longer than
it is assumed in standard models of NS cooling
(Neuh\"auser, Tr\"umper 1999, Popov et al. 2000b).

 Such high rate of recent supernova explosions in the solar vicinity
is in wonderful correspondence with recent results of
computer simulations of the Local Bubble formation
(Smith, Cox 2000, Ma\'iz-Apell\'aniz 2001). These authors argue, that
it is necessary to have at least 3-6 recent ($<10^7$ yrs) bursts
in the region close to the Sun and the youngest explosion should appeared
less than $<10^6$ years ago. Other data (even geophysical!) also
supports recent and close supernova explosions.

 As far as it is necessary to introduce magnetars in order
to explain periods of RINSs, we can also ask why relative fraction
of ultramagnetized NS is so high among RINSs.
The answer can be the following: due to MFD the crust of the INS
is heated, and it can stay hot for a longer time. This effect is especially
important for magnetars, that is why their fraction is so high.\\[2mm]     

{\it 3.3. ROSAT INSs: mixed population?}\\[1mm]
 
  As the nature of RINSs in unclear they can be not a one-type population,
but a mixture of coolers and accretors.

 It is very difficult to distinguish between cooling and accreting INSs.
The only case, about which most of scientists are sure that it is a cooler,
is RX J185635-3754. For this object parallax, proper motion 
and many other characteristics are known (see Pons et al. 2001
for the latest information about this object).

 Log N -- Log S calculations (Popov et al. 2000b) show, that it is nearly
impossible to explain significant part of RINSs by accretors with constant
field. Nobody made careful population synthesis for isolated accretors
with MFD. Most probably it is possible to find parameters for which
we can explain log N -- log S distribution with accretors with decayed
magnetic field, but the exact answer should come from observations. 
Future observations of $\dot p$ of RX J0720.4-3125, RBS1223 
(1RXS J130848.6+212708) and RX J0420.0-5022 
can provide direct evidence for field decay in INSs and help to 
distinguish between two interpretations.\\[2mm]

{\bf 4. Discussion}\\[1mm]

 Here we briefly discuss types of objects which 
are not described in details in previous sections.

 Pfahl and Rappaport (2001) suggested, that some of dim X-ray sources in
globular clusters can be old accreting INSs.
In our paper (Popov, Prokhorov 2001a) we checked this possibility
with a simple population synthesis model, and found, that results
of calculations were not in contradiction with observations.

 As far as no spin periods are observed for these objects one can suggest
two hypothesis: very long spin periods 
(see for example Popov, Prokhorov 2001b for discussion about spin periods
of old accreting INSs) or accretion onto
significant part of the NS's surface due to small magnetic field 
(Popov, Prokhorov 2001a). Relatively low temperatures of dim X-ray
sources suggest, that accretion proceeds not onto small polar caps,
corresponding to fields $<10^9$ G
(such values of magnetic field can be reached
in these extremely old objects due to MFD).
It can be an indication, that between two hypothesis the second one
is closer to reality. 

 It was suggested (Popov 1998), that a compact X-ray source inside 
the supernova remnant RCW~103 can be a relatively old accreting INS,
and the remnant itself was produced by an explosion of the secondary
component of a binary system. The object does not show any
modulation of radiation on short time scales (which can correspond
to spin period), 
but demonstrates flux variability on long time scale (years). 
It can be an indication of accretion onto very large polar caps,
corresponding to low magnetic field ($<10^9$ G). As far as such values are
not typical for most part of known young NSs (i.e. radiopulsars), one
can suggest, that magnetic field significantly decayed during the lifetime
of the NS.

 We conclude that different types of INSs show evidence for magnetic field
decay.\\[2mm]

{\it Acknowledgments.} We thank Monica Colpi, Roberto Turolla 
Aldo Treves and Denis Konenkov for discussions and
Universities of Como, Padova and Milan for hospitality.

SP thanks University of Como for financial support.

MP also thanks the organizing committee for support
of his participation in the conference.

This work was supported by grants of the
RFBR 01-02-06265 and 01-15(02)-99310.\\[3mm]


\indent
{\bf References\\[2mm]}
Colpi M., Geppert U., Page D.: 2000, {\it Ap.J.}, {\bf 529}, L29.\\
Colpi M., Turolla R., Zane S., Treves A.: 1998, {\it Ap.J.,} {\bf 501}, 252.\\
Colpi M., Possenti A., Popov S.B., Pizzolato F.: 2001,
in ``Physics of Neutron Star Interiors'', 
Eds. D. Blaschke, N.K. Glendenning, \& A. Sedrakian
(Springer--Verlag, Berlin), (astro-ph/0012394).\\
Duncan R.C.: 2001, astro-ph/0106041.\\
Duncan R.C., Thompson C.: 1992, {\it Ap.J.,} {\bf 392}, L9
Haberl F., Pietsch W., Motch C., Buckley D.A.H.: 1996, 
{\it Circ. IAU}, no. 6445.\\
Hurley K.: 2000, in "Gamma-ray Bursts", 5th Huntsville Symposium, 
Eds. R. M. Kippen, R. S. Mallozzi, G. J. Fishman,
published by American Institute of Physics, Melville, New York, p.763,
astro-ph/9912061.\\
Kaplan D. L., Kulkarni S. R., van Kerkwijk M. H., Rothschild R. E.,
Lingenfelter R. L.; Marsden D., Danner R., Murakami T.: 
2001, {\it Ap. J.,} {\bf 556}, 399.\\
Konenkov D.Yu.,  Popov S.B.: 1997, {\it Pisma Astron. Zh.,} {\bf 23}, 569.\\
Konenkov D.Yu., Geppert U.: 2001, {\it MNRAS}, {\bf 325}, 426.\\
Livio M., Xu C., Frank J.: 1998, {\it Ap.J.}, {\bf 492}, 298.\\
Ma\'iz-Apell\'aniz J.: 2001, astro-ph/0108472.\\
Mereghetti S.: 1999, astro-ph/9911252.\\
Neuh\"auser R.,  Tr\"umper J.E.: 1999, {\it As. Ap}, {\bf 343}, 151.\\
Pfahl  E., Rappaport S.: 2000, {\it Ap.J.}, {\bf 550}, 172,
astro-ph/0009212.\\
Pons J.A., Walter F.M., Lattimer J.M., Prakash M., Neuhaeuser R., Penghui
An: 2001, astro-ph/0107404.\\  
Popov S.B.: 1998, {\it Astron. Astroph. Trans.}, {\bf 17}, 35,
astro-ph/9708044.\\
Popov S.B., Colpi M., Treves A., Turolla R.,
Lipunov V.M., Prokhorov M.E.: 2000a, {\it Ap.J.,} {\bf 530}, 896.\\
Popov S.B., Colpi M., Prokhorov M.E., Treves A.,  Turolla R.:
 2000b, {\it Ap.J.}, {\bf 544}, L53.\\
Popov S.B.,  Prokhorov M.E.: 2000, {\it As. Ap.}, {\bf 357}, 164.\\
Popov S.B.,  Prokhorov M.E.: 2001a, {\it Astron. Astroph. Trans.}
(accepted), astro-ph/0102201.\\
Prokhorov M.E., Popov S.B., Khoperskov A.V.: 2001b, astro-ph/0108503.\\
Popov S.B.: 2001, astro-ph/0101031.\\
Smith R.K., Cox D.: 2000, {Ap. J. Supp.,} {\bf 134}, 283.\\
Treves A., Turolla R., Zane S., Colpi M.: 2000,
{\it Publ. Astron. Soc. Pac.}, {\bf 112}, 297.\\
Wang J.C.L.: 1997, {\it Ap.J.} {\bf 486}, L119.\\
Zampieri, L., Campana S., Turolla R., Chieregato M.,              
Falomo R., Fugazza D., Moretti A., Treves A.: 2000,
{\it As.Ap.} (accepted), astro-ph/0108456.\\
\vfill

\end{document}